\documentclass{jfm}
\usepackage{graphicx}
\usepackage{epstopdf, epsfig}

\shorttitle{Global instability of wing shock buffet}
\shortauthor{Sebastian Timme}

\title{Global instability of wing shock buffet}

\author{Sebastian Timme
  \corresp{\email{sebastian.timme@liverpool.ac.uk}}}

\affiliation{School of Engineering, The University of Liverpool,
Liverpool, England L69 3GH, UK}

\begin{document}

\maketitle

\begin{abstract}
Shock buffet on wings encountered in edge-of-the-envelope transonic flight remains an unresolved and disputed flow phenomenon, challenging both fundamental fluid mechanics and applied aircraft aerodynamics. The question of global instability leading to flow unsteadiness is addressed herein. It is shown for the first time that incipient three-dimensional shock buffet is governed by the dynamics of a single unstable linear eigenmode with its spatial structure describing previously reported buffet cells. An inner-outer Krylov approach is proposed to solve the arising large-scale eigenvalue problem iteratively using shift-and-invert spectral transformation and sparse iterative linear solver. The established numerical strategy with an industrial flow solver means that a practical non-canonical test case at high-$\Rey$ flow condition can be investigated. A modern wing design with publicly available geometry and experimental data for code validation is studied. The numerical findings can be exploited for routes to flow control and model reduction.
\end{abstract}


\section{Background}

Shock buffet on wings is an undesirable phenomenon limiting the flight envelope at high Mach numbers and load factors. Its study is critical for commercial transonic air transport. The term commonly refers to an aerodynamic instability with self-sustained shock-wave oscillations and intermittent boundary layer separation. Whereas aerofoil buffet in turbulent flow is characterised by large chord-wise shock excursions at dominant Strouhal numbers of 0.06 to 0.07, wing buffet typically comes with lower amplitude shock motions and is more broadband with up to an order of magnitude higher frequencies 
\citep{dandois2016}. A span-wise outboard propagation of `buffet cells' (a term coined by~\citet{iovnovich2015}), which is believed to constitute the instability, has been reported both in experimental and numerical studies~\citep{lawson2016,koike2016,sartor2016}. A span-wise inboard propagation has also been identified experimentally at lower frequencies~\citep{dandois2016,masini2017}.  
\citet{Timme2016} observed resonant flow due to forced wing vibration in the same lower frequency range, in addition to distinct flow responses around typical shock buffet frequencies on wings. 
While the flow unsteadiness is self-excited, not requiring structural vibration itself, resulting 
aerodynamic loads excite the wing structure (called buffeting) thus deteriorating passenger comfort, flight control and performance and the fatigue life. Certification specifications stipulate that an aircraft must be free of any vibration and buffeting in cruising flight with a margin of 0.3g to the buffet onset boundary. 

Shock buffet characteristics on aerofoils and wings are distinct, and despite more than half a century of research an unequivocally agreed physical interpretation is still debated \citet{giannelis2017}. 
An important theoretical/numerical advance was the \cite{crouch2007,crouch2009} discovery of a global (asymptotic, modal, absolute) instability leading to aerofoil buffet, using Reynolds-averaged Navier--Stokes (RANS) aerodynamics in a base-flow scenario. The interested reader is referred to the excellent reviews by~\citet{sipp2010} and \citet{theofilis2011} for a reflection on the various terms denoting such oscillator-type flow instability resulting from a Hopf bifurcation. A base-flow approach essentially refers to linearising both the RANS equations and a turbulence model around an equilibrium point (i.e.~a steady-state solution)~\citep{mettot2014b}. Even though Crouch's description of the instability somewhat differs from the widely discussed model by~\citet{lee1990}, the two models both rely on an acoustic feedback mechanism and involvement of the trailing edge, an observation which is also supported by eddy-resolving simulation~\citep{deck2005} 
and experiment~\citep{Feldhusen-Hoffmann2017}. \citet{Sartor2015} additionally identified a convective Kelvin--Helmholtz-type instability via optimal forcing responses using the resolvent approach. In the three-dimensional case, \citet{iovnovich2015} pursued the categorisation of the three-dimensionality of wing buffet, using time-marching unsteady RANS, by progressively building up the geometric complexity. 
Scale-resolving detached eddy simulation on the other hand has been applied for wing buffet flow by \citet{brunet2008aiaa} and \citet{sartor2016}, supporting the above mentioned span-wise propagation of `buffet cells'. At the same time, industrial practice mostly relies on steady RANS analysis e.g. with the `$\Delta\alpha=0.1^\circ$ offset' method \citep{kenway2017} to decide on shock buffet onset.

In recent years, modal descriptions of shock buffet on wings have been pursued intensively. \citet{ohmichi2017} applied modal identification techniques, specifically proper orthogonal decomposition and dynamic mode decomposition, to discern dominant modal aerodynamic behaviour from solution snapshots well beyond buffet onset. Focussing instead on the discretised RANS (plus turbulence model) operator directly, global mode computation on a case with three inhomogeneous spatial dimensions has first been accomplished in pre-buffet conditions by \citet{Timme2016}. 
Although not the first reported three-dimensional stability analysis (see for example~\citet{theofilis2011} for a short, yet quickly growing list), the work focussed exclusively on geometric non-canonical complexity and flow parameters, specifically high $\Rey$, relevant to an aircraft wing. 
A conclusive identification of the sought unstable global mode, with the chosen numerical approach, failed due to non-converging base flow beyond suspected buffet onset, and this would require, for instance, a matrix-free, time-stepping iterative tool for modal analysis~(see for example~\citet{eriksson1985,mack2010}).

Previous aerofoil buffet studies using global stability analysis applied sparse direct linear equation solvers with a full factorisation of the coefficient matrix. The bottleneck is the excessive memory requirement that has already been observed for simple aerofoil cases
~\citep{iorio2014}. This renders direct methods infeasible for truly three-dimensional cases, when solving linear systems arising from a shift-and-invert approach, used for instance in the implicitly restarted Arnoldi method~\citep{sorensen1992}. A viable alternative is to use sparse iterative linear equation solvers, and the generalized minimal residual method~\citep{saad1986} has become standard practice. Trading memory requirements for computing time, such iterative methods often stall for very stiff problems, found in transonic turbulent flow near buffet onset and exacerbated by the nearly singular shift-and-invert preconditioned matrix eigenvalue problem. \citet{Timme2016} opted for a Krylov method with deflated restarting to make their tools robust.

An important, and currently missing, link in the fundamental understanding of the very basics of three-dimensional shock buffet, analogous to the seminal aerofoil work by \citet{crouch2007,crouch2009}, is confirmation of the existence of an unstable global mode, or even multiple modes indeed. This is the central question to be addressed in this work. Section~\ref{sec:numerics} introduces the numerical approach, followed by Section~\ref{sec:testcase} outlining the chosen test case and some basic validation of the simulations. Details of the unstable global mode describing the incipient shock buffet instability are presented in Section~\ref{sec:results}.


\section{Numerical approach}\label{sec:numerics}

The aerodynamics are simulated herein using the industry-grade DLR-TAU software package~\citep{Schwamborn2006}. The compressible RANS equations are solved with a second-order finite-volume discretisation. Turbulent closure is achieved with the negative version of the Spalart--Allmaras model~\citep{Allmaras2012}. \citet{langer2014} provides a good account of the code's spatial discretisation. Specifically, inviscid fluxes are evaluated with a central scheme with matrix artificial dissipation, and gradients of flow variables for viscous fluxes and source terms are computed using the Green--Gauss theorem. Steady base-flow solutions are obtained using the backward Euler method with lower-upper Symmetric--Gauss--Seidel iterations and local time-stepping. Convergence is further accelerated through the use of a 4w multigrid scheme. All steady simulations herein converged to at least 11 orders of magnitude in the density residual norm. 

For time-marching unsteady simulations, the governing equations are integrated in time using the second-order backward differentiation formula with subiterations at each time step. A Cauchy convergence criterion with a relative error tolerance of $10^{-8}$ on the drag coefficient is applied on the subiteration level in addition to monitoring the normalised density residual norm ($10^{-3}$). A minimum of 50 subiterations per physical time step is always performed for the simulations presented. Criteria on iterations and the chosen time-step size ($\Delta t=1\,\mu s$) follow previous studies~\citep{sartor2016}.

Global stability analysis concerns the asymptotic time evolution of infinitesimal perturbations $\varepsilon\widetilde{\boldsymbol{u}}$ to a base flow $\bar{\boldsymbol{u}}$, with ${\boldsymbol{u}}$ 
containing the conservative variables of the RANS equations plus turbulence model at each grid point and $\varepsilon\ll 1$. Interest is in solutions of the form $\widetilde{\boldsymbol{u}}=\widehat{\boldsymbol{u}}e^{\lambda t}$ where $\widehat{\boldsymbol{u}}$ is the three-dimensional spatial structure of the eigenmode (i.e.~right/direct eigenvector) and $\lambda=\sigma+i\omega$ describes its temporal behaviour (i.e.~eigenvalue) with $\sigma$ as the growth rate and $\omega$ as the angular frequency. Linearisation of the spatial discretisation of the non-linear governing equations (including turbulence model), and substitution of the solution, leads to an algebraic system of equations,
\begin{equation}
\mathsfbi{J}\widehat{\boldsymbol{u}}=\lambda\widehat{\boldsymbol{u}}
\label{eq:eig}
\end{equation}
where 
$\mathsfbi{J}$ is the fluid Jacobian matrix (i.e.~the linearisation). For eigenmode computations, the implicitly restarted Arnoldi method proposed by~\citet{sorensen1992} and implemented in the ARPACK library~\citep{arpack,Maschhoff1996} has been coupled with the linear harmonic incarnation of the chosen flow solver. Since this Arnoldi method has been explained many times in the literature~(see for example~\citet{mack2010}), it is only summarised briefly here. In essence, Arnoldi's method is used to approximate a few eigenmodes of $\mathsfbi{J}$. The approximation of eigenmodes improves with the number of Krylov vectors and restarting is applied in practice. A polynomial approximation of the restart vector is key to the method. For detail refer to \citet{sorensen1992}. Shift-and-invert spectral transformation is applied to converge to wanted parts of the eigenspectrum, and critical is therefore the robust solution of many linear systems of equations.

\begin{table}
	\begin{center}
		\def~{\hphantom{0}}
		\begin{tabular}{lc}
			Parameter  & Value  \\[3pt]
			Maximum number of eigenmodes per shift & ~20 \\
			Maximum number of outer iterations & ~~3 \\
			Size of Krylov space for outer iterations   & 100\\
			Convergence criterion on outer iterations   & 10$^{-6}$\\
			Size of Krylov space for inner iterations   & 120\\
			Number of deflation vectors for inner iterations   & ~20\\
			Convergence criterion on inner iterations   & 10$^{-7}$\\
		\end{tabular}
		\caption{Overview of numerical settings for eigenvalue solver}
		\label{tab:num_para}
	\end{center}
\end{table}

The linearised frequency-domain flow solver follows a first-discretise-then-linearise, matrix-forming philosophy with a hand-differentiated Jacobian matrix~$\mathsfbi{J}$. Implementation details in DLR-TAU are provided by~\citet{Dwight2006} and \citet{Thormann2013}. Pivotal to solve arising linear systems is the generalised conjugate residual algorithm with deflated restarting~\citep{parks2006,Xu2016}. To offer the essential insight into the chosen Krylov method, a first basis of Arnoldi vectors is always computed using the standard generalized minimum residual algorithm~\citep{saad1986}. Whereas basic restarted Krylov solvers usually discard all available information during restart, only to rebuild the entire subspace from scratch again, the chosen advanced iterative solver aims to retain key information which is found by ranking the interior eigenvalues, approximated by the Hessenberg matrix. This often results in a more robustly converging iteration combined with lower memory usage due to a smaller required Krylov subspace. For preconditioning, a local block incomplete lower-upper factorisation of the shifted Jacobian matrix with zero level of fill-in is selected~\citep{mccracken2013}. 

Numerical settings of the inner-outer Krylov approach, i.e.~the inner sparse iterative linear equation solver and the outer iterative eigenvalue solver, are provided in table~\ref{tab:num_para}. 
The strength of the approach lies in its numerical algorithms and not in the ultimate of brute-force high-performance computing. All jobs ran on two nodes, each having twin Skylake 6138 processors, $40$ hardware cores and 384 GB of memory. Each linear solution takes less than an hour for the test case with nearly 37~million degrees-of-freedom.

\section{NASA Common Research Model}\label{sec:testcase}

The NASA Common Research Model is a generic commercial wide-body aircraft configuration. It was developed to publicly make available a modern supercritical wing geometry together with state-of-the-art experimental data, enabling code validation, with tests completed in several transonic wind tunnel facilities~\citep{vassberg2008}. The wing was designed to have an aspect ratio of 9, a taper ratio of 0.275 and a $35^\circ$ quarter-chord sweep angle. The mean aerodynamic chord of the wind tunnel model is $0.189~\textrm{m}$ with a span and reference area of $1.586~\textrm{m}$ and $0.280~\textrm{m}^2$, respectively. All design details including aerofoil data can be found in cited reference. The present study analysed the wing-body-tail variant with $0^\circ$ tail setting angle, discarding pylon and nacelle.

The computational mesh was generated using the SOLAR mesh generator following accepted industrial practice and has about $6.2$ million points including $167\,190$ on solid walls for the half model used (excluding the blade sting mounting system). A viscous wall spacing of $y^+<0.5$ is ensured. The hemispherical farfield boundary is located approximately 100 semi-spans from the body, while a symmetry boundary condition is applied along the centre plane. Flow parameters are chosen for runs 153/182 of the test campaign in the European Transonic Windtunnel (ETW), with the full data set available at \textit{http://www.eswirp.eu/ETW-TNA-Dissemination.html}. Specifically, Mach number is $M=0.85$ and Reynolds number is $\Rey=5$~million per reference chord. No transition fixing was used in the simulations, contrary to experiments at this $\Rey$, and fully turbulent flow is assumed. Run~182 measured the static deformation of the flexible wing at seven angles of attack and the computational mesh was deformed accordingly, a functionality readily available in the chosen flow solver, to improve numerical predictions~\citep{tinoco2017}. Wind tunnel force measurements shown below have been corrected for wall interference and include a correction due to buoyancy effects of the mounting system.

\begin{figure}[t]
	\centering
	\includegraphics[width=1\textwidth,trim=0 333 0 0 ]{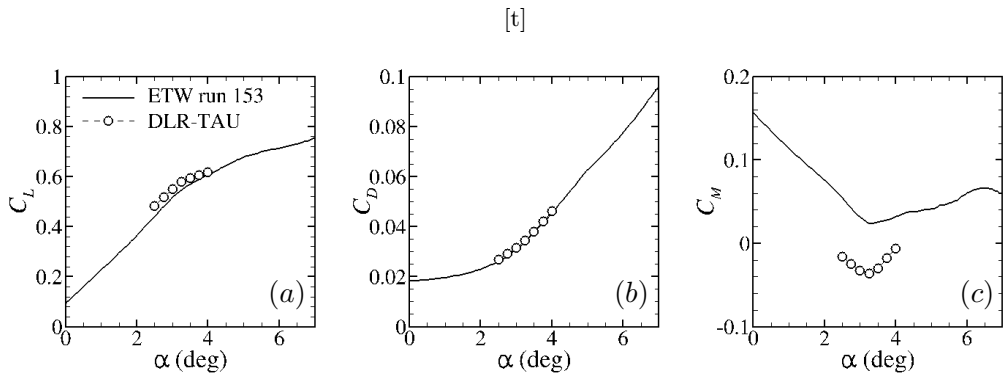}
	\put(-285,25){\large{$(a)$}}
	\put(-155,25){\large{$(b)$}}
	\put( -25,25){\large{$(c)$}}
	\caption{Aerodynamic loads at $M=0.85$ and $\Rey=5\,$ million comparing experiment and simulation for (\textit{a}) lift coefficient, (\textit{b}) drag coefficient and (\textit{c}) pitching moment coefficient.}
	\label{fig:loads}
\end{figure}
\begin{figure}[t]
	\centering
	\includegraphics[width=\textwidth]{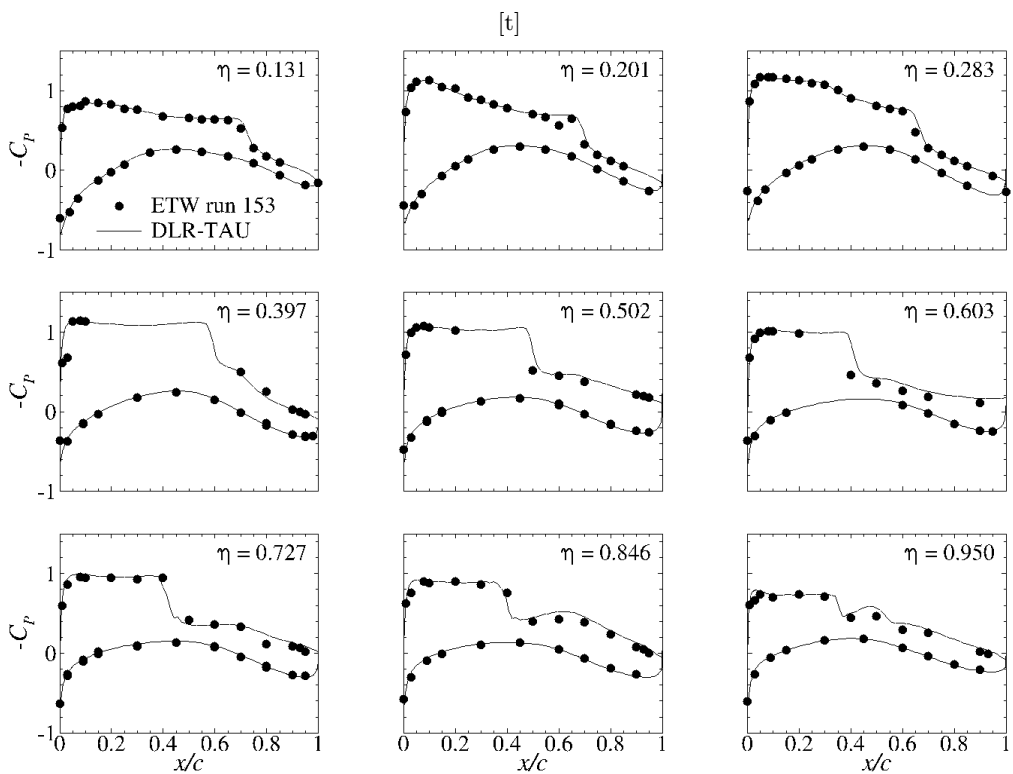}
	\caption{Surface pressure coefficient ${C}_p$ at $M=0.85$, $\Rey=5\,$ million and $\alpha=3.75^\circ$ comparing experiment and simulation along nine non-dimensional span-wise stations $\eta$.}
	\label{fig:surfacepressure}
\end{figure}

Figures~\ref{fig:loads} and \ref{fig:surfacepressure} show a basic validation of the simulations reported herein. Aerodynamic coefficients at seven angles of attack, given in figure~\ref{fig:loads}, suggest a fairly good agreement between the RANS simulations and wind tunnel data, when compared to other numerical predictions~\citep[see for example][]{tinoco2017}. The break in lift and moment curves occurs at about $\alpha=3.3^\circ$ angle of attack, similar to wind tunnel data. The clear offset in moment coefficient, reported elsewhere, too, is not fully understood but could result from the partial correction applied for the mounting system. Surface pressure distributions in figure~\ref{fig:surfacepressure} at nine span-wise stations assert these favourable conclusions. More in-depth validation is out-of-scope and the reader is referred to published collaborative studies. Subsequent focus is on angles of attack of $\alpha=3.50^\circ$ and $3.75^\circ$.

\section{Global linear instability results}\label{sec:results}

\begin{figure}
	\centering
	\includegraphics[width=0.495\textwidth, trim= 0 20 0 0]{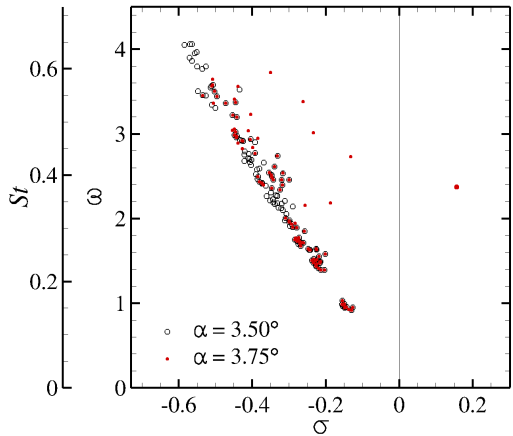}
	\caption{Computed eigenvalues for angles of attack of $\alpha=3.50^{\circ}$ and $3.75^{\circ}$.}
	\label{fig:eigenvalues}
\end{figure}

\begin{figure}
	\centering
	\includegraphics[width=0.495\textwidth, trim= 0 20 0 0]{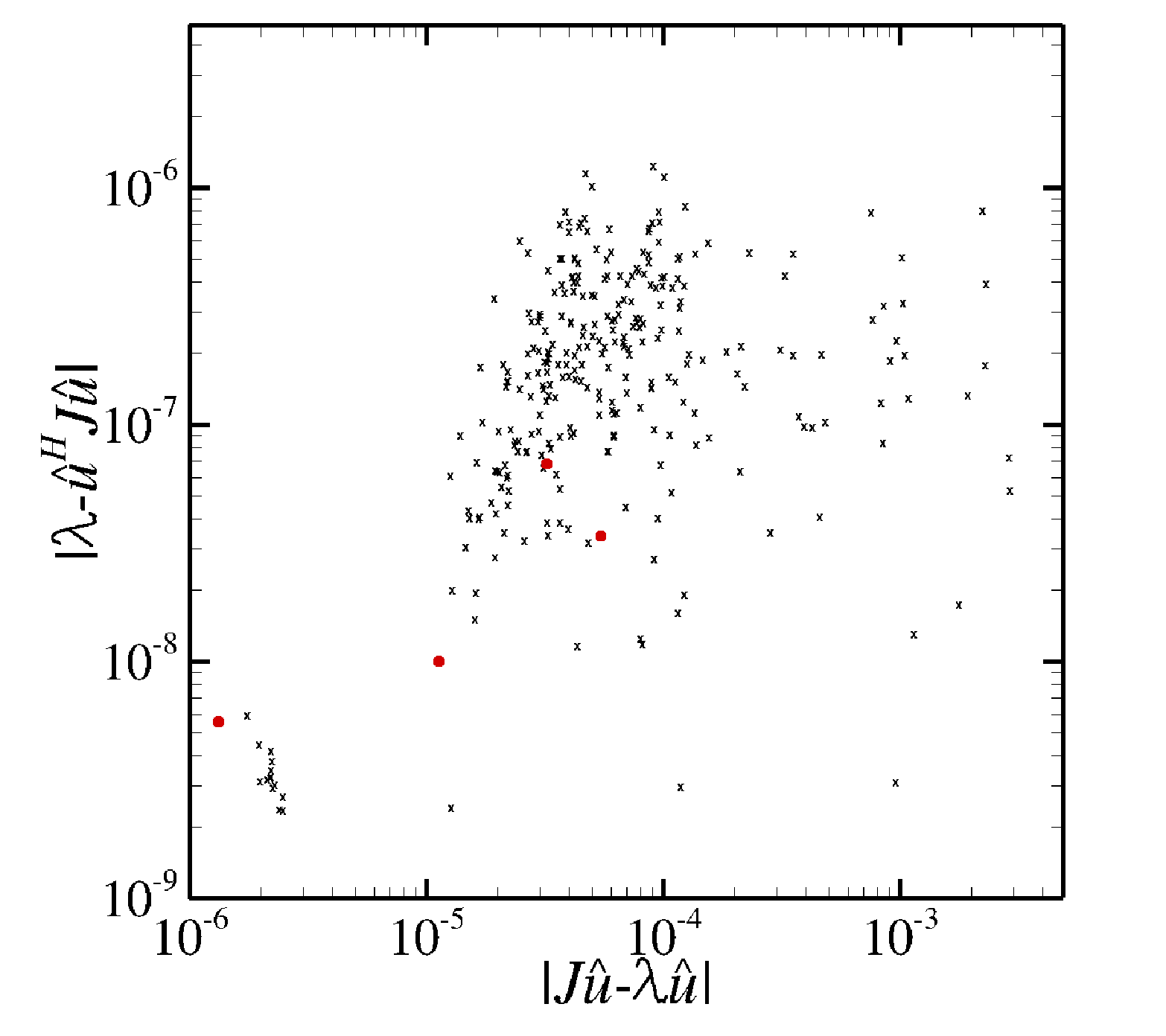}
	\includegraphics[width=0.495\textwidth, trim= 0 20 0 0]{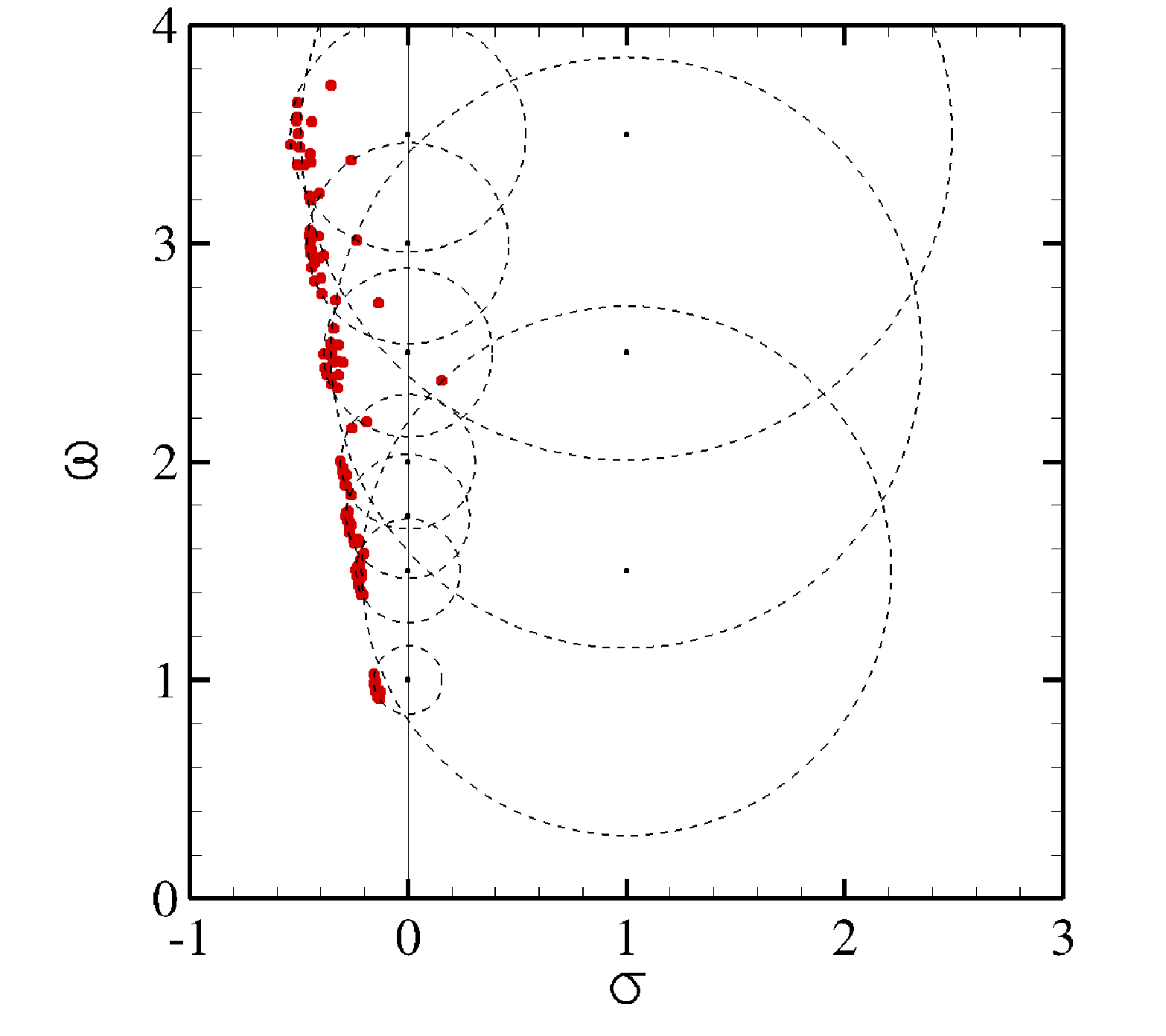}
	\put(-230,145){\large{$(a)$}}
	\put( -35,145){\large{$(b)$}}		
	\caption{Sanity checks on convergence and coverage of eigenmode computations showing (\textit{a})~norm of Rayleigh quotient error $\|\lambda-\widehat{\boldsymbol{u}}^H\mathsfbi{J}\widehat{\boldsymbol{u}}\|$ vs.~residual norm of eigenvalue problem $\|\mathsfbi{J}\widehat{\boldsymbol{u}}-\lambda\widehat{\boldsymbol{u}}\|$ for all computations and (\textit{b}) intersection and union of solutions based on individual shifts at $\alpha=3.75^\circ$. Red dots in (\textit{a}) are the multiple solutions of the unstable mode.}
	\label{fig:sanity}
\end{figure}

Details of the three-dimensional global stability computations are discussed in the following. All results are presented in non-dimensional units unless otherwise stated. Figure~\ref{fig:eigenvalues} shows the computed eigenvalues for the two angles of attack where buffet onset is expected. Several shifts were distributed along the imaginary axis in addition to a few shifts with positive growth rate, enabling a wider search with reduced convergence rate of the shift-and-invert spectral transformation. The lower angle of attack of $\alpha=3.50^\circ$ describes a subcritical flow. The higher angle at $\alpha=3.75^\circ$ constitutes a shock buffet condition. Besides a few eigenvalues with increased growth rate for Strouhal numbers $St=\omega/(2\pi)$ between $0.35$ and $0.6$, which show a similar pattern as presented by~\citet{Timme2016} for an older-generation wing design and could explain the broadband-frequency nature of wing shock buffet beyond onset conditions, a single unstable global mode with a dimensionless growth rate of $\sigma=0.156$ and an angular frequency of $\omega=2.371$ (corresponding to a Strouhal number of $St=0.377$) is found.

\begin{figure}
	\centering
	\includegraphics[width=\textwidth,trim=0 295 0 0 ]{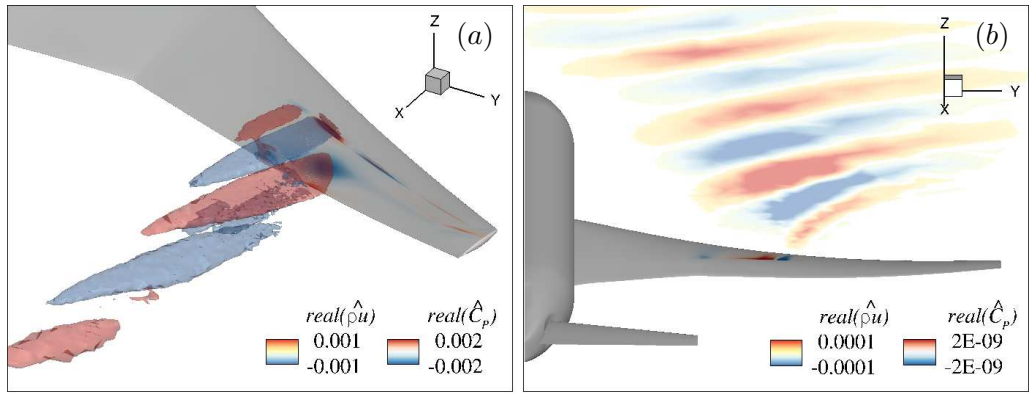}
	\put(-215,130){\large{$(a)$}}
	\put( -20,130){\large{$(b)$}}
	\caption{Spatial structure of unit-length unstable eigenmodes showing real part of both x-momentum $\widehat{\rho u}$ and surface pressure coefficient $\widehat{C}_p$; (\textit{a}) right/direct mode with spatial iso-surface of $\widehat{\rho u}$, (\textit{b}) left/adjoint mode with contours of $\widehat{\rho u}$ in $xy$-plane aligned with wing.}
	\label{fig:field}
\end{figure}

Figure~\ref{fig:sanity} establishes numerical credibility of presented data. The norms of the Rayleigh quotient error $(\lambda-\widehat{\boldsymbol{u}}\,\!^H\mathsfbi{J}\widehat{\boldsymbol{u}})$ and the residual of the eigenvalue problem $(\mathsfbi{J}\widehat{\boldsymbol{u}}-\lambda\widehat{\boldsymbol{u}})$ are shown for all computations in figure~\ref{fig:sanity}(\textit{a}). The residual norm does not agree with the convergence criterion set on the outer iteration (cf.~table~\ref{tab:num_para}), which is due to ARPACK's approach of assessing convergence based on the Hessenberg matrix~\citep{arpack}. The isolated group in the lower left corner resulted from solving an adjoint problem corresponding to equation~(\ref{eq:eig}), specifically  $\mathsfbi{J}^T\widehat{\boldsymbol{v}}={\lambda}^\ast\widehat{\boldsymbol{v}}$ with $\widehat{\boldsymbol{v}}$ as the left/adjoint eigenvector and $\lambda^\ast=\sigma-i\omega$ as complex conjugate eigenvalue, keeping all other parameters the same. Such favourable convergence should be exploited in the future. Multiple converged solutions for the unstable mode due to overlapping search regions are highlighted by red dots in figure~\ref{fig:sanity}(\textit{a}). In figure~\ref{fig:sanity}(\textit{b}), coverage of the relevant part of the eigenspectrum (where `relevant' relates to engineering judgement) is sufficient and has a large overlap. The radius of a circle describes the greatest distance between a shift (black dots in the figure) and any of its converged eigenvalues. While such an approach to find rightmost eigenvalues appears naive, mathematically rigorous algorithms to compute those eigenvalues directly (see for example~\citet{elman2012}) do not seem feasible for the problem size, as yet.

\begin{figure}
	\centering
	\includegraphics[width=\textwidth]{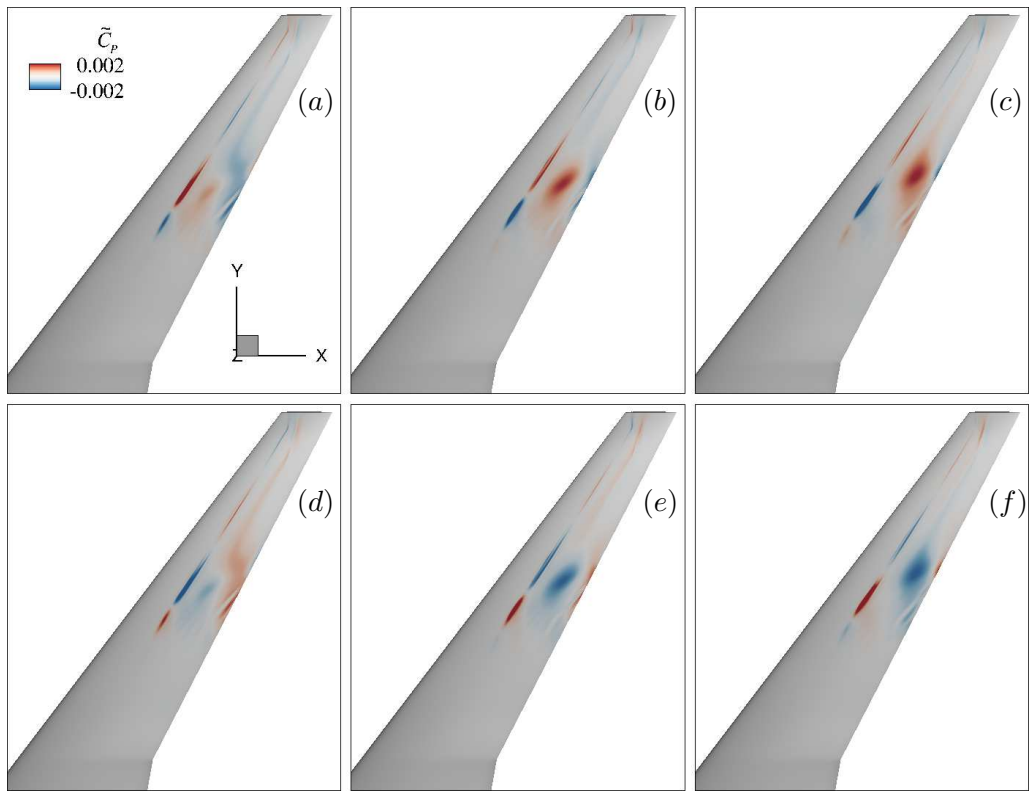}
	\put(-275,255){\large{$(a)$}}
	\put(-145,255){\large{$(b)$}}
	\put( -15,255){\large{$(c)$}}
	\put(-275,105){\large{$(d)$}}
	\put(-145,105){\large{$(e)$}}
	\put( -15,105){\large{$(f)$}}        		
	\caption{Time evolution of unit-length unstable mode showing disturbance in surface pressure coefficient $\widetilde{C}_p(x,y,z,t)$ at six time steps during oscillation cycle with period $T$; (\textit{a}) $t/T=0=1$, (\textit{b}) $t/T=1/6$, (\textit{c}) $t/T=2/6$, (\textit{d}) $t/T=4/6$, (\textit{e}) $t/T=5/6$, (\textit{f}) $t/T=5/6$.}
	\label{fig:surface}
\end{figure}

The spatial structure of the right and left unstable modes is presented in figures~\ref{fig:field} and~\ref{fig:surface}, visualising buffet cells. The term `buffet cell' refers to a local flow arrangement of a ripple along the span-wise shock wave combined with a pulsating recirculation bubble. In figure~\ref{fig:field} only the real part is shown since the imaginary part is $90^\circ$ out-of-phase to enable the downstream convection of structures~\citep{crouch2007,sipp2010}. A certain similarity with the two-dimensional aerofoil analysis by~\citet{Sartor2015}, who, for example, emphasised opposite signs in the x-momentum $\widehat{\rho u}$ within the shock wave and separation region, is striking. While the shock moves downstream, the bubble contracts, and vice versa. A complicating factor is the added three-dimensionality with propagation not only chord-wise but also span-wise, which can be noticed in the lag of x-momentum structures downstream in the wake region. Visual inspection of the other eigenmodes with increased growth rate (cf.~figure~\ref{fig:eigenvalues} for Strouhal numbers around the critical value) indicates phase-shifted, yet very similar spatial structures (not shown herein). Visualising and interpreting the left eigenvector is less intuitive. Its spatial structure emphasises the region where a harmonic forcing would optimally trigger the buffet instability~\citep{sipp2010}. Specifically in figure~\ref{fig:field}(b), the unstable adjoint mode with strong upstream support to allow a downstream convection of information forms a conical shape opening in the upstream direction. The figure presents an $xy$-plane aligned with the wing. Figure~\ref{fig:surface} shows the time evolution of the surface pressure coefficient reconstructed from the unit-length right eigenmode using $\widetilde{C}_p(x,y,z,t)=\Real(\widehat{C}_p(x,y,z)\, e^{i\omega t})$ at six time steps during one period of oscillation (disregarding the mode amplification). The repeated span-wise outboard (and chord-wise downstream) propagation of buffet cells, as previously suggested (see for example~\citet{sartor2016}), is corroborated. 

\begin{figure}
	\centering
	\includegraphics[width=\textwidth,trim=0 360 0 0 ]{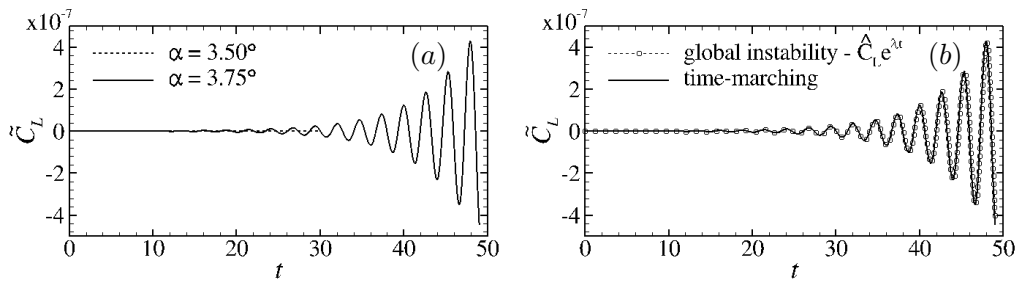}
	\put(-230,80){\large{$(a)$}}
	\put(-35,80){\large{$(b)$}}
	\caption{Results of lift coefficient $\widetilde{C}_L$ showing (\textit{a}) time-marching unsteady RANS bracketing buffet onset between $\alpha=3.50^\circ$ and $3.75^\circ$ and (\textit{b}) time-marching unsteady RANS compared with unsteady signal based on unstable eigenmode at $\alpha=3.75^\circ$.}
	\label{fig:timedomain}
\end{figure}

To further support the findings, conventional time-marching unsteady RANS results are presented in figure~\ref{fig:timedomain}. Time histories of the lift coefficient in figure~\ref{fig:timedomain}(\textit{a}) locate the onset of buffet instability between $\alpha=3.50^\circ$ and $3.75^\circ$. No attempt is made to refine this bracket. Figure~\ref{fig:timedomain}(b) compares the lift coefficient of time-marching unsteady RANS at $\alpha=3.75^\circ$ with the lift coefficient calculated from the unstable eigenmode using $\widetilde{C}_L(t)=\widehat{C}_Le^{\lambda t}$. Specifically, the relation $\widehat{C}_L=\partial{C}_L/\partial{\boldsymbol{u}}\cdot \widehat{\boldsymbol{u}}$ holds. Since the eigenvector~$\widehat{\boldsymbol{u}}$ is scaled to unit length by default, its magnitude is adjusted to match time-domain results. Agreement between conventional and global stability analysis is excellent.

\section{Conclusions}

Eigenmodes of a three-dimensional test case, specifically an aircraft wing in high-$\Rey$ turbulent and transonic flow, have been computed. A matrix-forming iterative scheme of an inner-outer Krylov structure succeeds in identifying an absolute instability linked to shock buffet on a wing for the first time. These fundamental results suggest that incipient shock buffet is dominated by the dynamics of a single unstable linear eigenmode.

The numerical findings are surprising in light of the often-used broadband-frequency explanation of three-dimensional buffet and have far-reaching implications, going beyond a mere better understanding of edge-of-the-envelope flow physics. This study will inform routes to buffet control via eigenvalue sensitivity and when attempting to establish shock buffet prediction tools for routine industrial analysis, such as reduced order models based on modal decomposition and projection. The adapted industrial flow solver paves the way to exploit concepts, established in fundamental fluid mechanics on mostly canonical test cases, in an applied and practical setting. It is expected that higher-fidelity eddy-resolving simulations, to overcome well-known inherent issues of turbulence modelling, will result in the same low frequency buffet mode. With an absolute instability confirmed, the role of convective mechanisms in shock buffet flow physics remains to be scrutinised.

\bibliographystyle{jfm}
\bibliography{arxiv-version}

\end{document}